# A Tractable Approach to Finding Closest Truncated-commute-time Neighbors in Large Graphs

**Purnamrita Sarkar**
Machine Learning Department
Carnegie Mellon University
Pittsburgh, PA 15213

**Andrew W. Moore**
Google Inc.
Pittsburgh, PA 15213

## Abstract

Recently there has been much interest in graph-based learning, with applications in collaborative filtering for recommender networks, link prediction for social networks and fraud detection. These networks can consist of millions of entities, and so it is very important to develop highly efficient techniques. We are especially interested in accelerating random walk approaches to compute some very interesting proximity measures of these kinds of graphs. These measures have been shown to do well empirically (Liben-Nowell & Kleinberg, 2003; Brand, 2005). We introduce a truncated variation on a well-known measure, namely *commute times* arising from random walks on graphs. We present a very novel algorithm to compute all *interesting* pairs of approximate nearest neighbors in truncated commute times, without computing it between all pairs. We show results on both simulated and real graphs of size up to 100,000 entities, which indicate near-linear scaling in computation time.

## 1 Introduction

The main aim of link-prediction and collaborative-filtering is to answer the question, for any selected node: which other nodes in the graph are *nearest* to this node? It is important to have a useful proximity measure for this purpose and very desirable for near nodes to be computed quickly. Ideally we will want this similarity metric to capture the graph structure. For example if two nodes have many common neighbors, they are highly similar. On the other hand, imagine a graph with two distinct connected components, connected by a single link. The two nodes connecting the components are neighbors in the graph but belong to different clusters, and are expected to be less similar than their neighbors in the same component.

A rather intuitive way of capturing this is the expected path length from one node to another during a random walk. This measure is called the hitting time (Aldous & Fill, 2001), and it tells us how long it will take on an average to hit the destination node from a source node. Note however that this measure is not guaranteed to be symmetric. Hence it's easier to work with the round trip time, i.e. the *commute time* between two nodes. These measures are inherently robust to noise and exploit the information encoded in the graph structure. They are widely used for unsupervised and semi-supervised learning (Zhu et al., 2003).

In spite of being very successful for collaborative filtering (Brand, 2005), dimensionality reduction (Saerens et al., 2004) or image segmentation problems (Qiu & Hancock, 2005), the main drawback of these measures is the computational load. In previous work authors have used approximate sparse matrix inversion techniques (Brand, 2005) or subspace approximations using principal eigenvectors (Saerens et al., 2004) to avoid $O(n^3)$ computation. However these techniques do not scale to graphs with more than few thousands of nodes. Brand (2005) uses clever iterative approaches to avoid this bottleneck, but some quantities derived from the random walk are loosely approximated.

In this paper we introduce a truncated variant of random walks in order to exploit local structure in graphs, and devise a novel algorithm to compute all interesting pairs of near-neighbors under this truncated version of commute times without having to compute the entire commute times matrix. Results on both simulated and real networks of size up-to 100,000 (on a single processor) indicate near-linear scaling in computation time.

## 2 Random walks on undirected graphs

In this section we introduce some basic concepts of random walks on undirected graphs. An undirected graph $G$ has a set of vertices $V$ numbered 1 to $n$, and a set of edges $E$. An edge is a pair $(i,j)$ where $i,j \in V$, and $(i,j) \in E \Leftrightarrow (j,i) \in E$. The adjacency matrix $W$ of $G$ is an $n \times n$ matrix of real values, where $W_{ij}$ denotes the weight on edge $i,j$, and is zero if the edge does not exist. For undirected graphs $W$ is symmetric. $D$ is an $n \times n$ diagonal matrix, where $D_{ii} = \sum_j W_{ij}$. We will denote $D_{ii}$ by the degree $d(i)$ of node $i$ from now on. The Laplacian $L$ of $G$ is defined as $D - W$.

Consider a random walk (Lovasz, 1996) on $G$ such that, if at step $t$ we are in vertex $v_t = i$, then in the next step we move to a neighbor of $i$ with probability proportional to the weight $w_{ij}$ of the link, i.e. $\frac{w_{ij}}{D_{ii}}$. Clearly the sequence of nodes $v_0, v_1, ..., v_t$ form a markov chain. We have $P_t(i) = Pr(v_t = i)$. $P = p_{ij}, i, j \in V$ denotes the transition probability matrix of this markov chain, so that $p_{ij} = w_{ij}/D_{ii}$ if $(i,j) \in E$ and zero otherwise.

In a random walk, if node $v_0$ is chosen from a distribution $P_0$, then the distributions $P_0, P_1, ..$ are in general different from one another. However if $P_0 = P_1$, then we say that $P_0$ is the stationary distribution for the graph. It can be shown that for a graph $G$ the stationary distribution is given by $\pi(v) = \frac{d(v)}{V(G)}$, where $V(G) = \sum_v d(v)$, denotes the volume of the graph.

We now introduce two main proximity measures derived from random walks, namely the hitting and the commute time (Aldous & Fill, 2001).

**Hitting time $H$:** The hitting time between nodes $i$ and $j$ is defined as the expected number of steps in a random walk starting from $i$ before node $j$ is visited for the first time. $H = h_{ij}, i, j \in V$ is an $n \times n$ matrix. Recursively $h_{ij}$ can be written as $h_{ij} = 1 + \sum_k p_{ik} h_{kj}$ if $i \neq j$ and zero otherwise.

Hitting times are unsymmetric, however they follow the triangle inequality (Lovasz, 1996).

**Commute time $C$:** Commute time between a pair of nodes is defined as $c_{ij} = h_{ij} + h_{ji}$ and is symmetric. $c_{ij}$ can also be defined as

$$c_{ij} = V(G)(l_{ii}^+ + l_{jj}^+ - 2l_{ij}^+)$$
$$= V(G)(e_i - e_j)^T L^+ (e_i - e_j) \quad (1)$$

Since each row of the graph Laplacian sums to zero, it has an eigenvector of all ones, with eigen-value zero. It is common practice (Saerens et al., 2004) to shift the zero-eigenvalue by subtracting the all ones matrix, inverting the resulting matrix, and then adding the all ones matrix back again, i.e. $L^+ = (L - \frac{1}{n}\mathbf{1}\mathbf{1}^T)^{-1} + \frac{1}{n}\mathbf{1}\mathbf{1}^T$.

The pseudo-inverse of $L$, i.e. $L^+$, can be viewed as a kernel (Smola & Kondor, 2003) which maps each vertex of a graph to a Euclidian space $i \mapsto x_i$. Pairwise commute time can be expressed as the squared Euclidian distance in the transformed space (equation (1)), where $x_i = (L^+)^{\frac{1}{2}} e_i$. Let us now look at the entries of $L^+$ in terms of these position vectors. The $ij$-th entry $l_{ij}^+ = x_i^T x_j$, denotes the dot-product between the position vectors of vertex $i$ and $j$. The diagonal element $l_{ii}^+$ denotes the squared length of the position vector of $i$. The cosine similarity (Brand, 2005) between two nodes is defined as $l_{ij}^+/\sqrt{l_{ii}^+ l_{jj}^+}$. We discuss different ways of computing the above quantities in Section 3.

**Robustness towards noise:** Hitting and commute times are robust to noise. Doyle and Snell (1984) view an undirected unweighted graph as an electrical network with every edge replaced by an unit weight resistor. In Chandra et al. (1989) the authors show that the commute time between two nodes $s, t$ in a graph is proportional to the effective resistance between $s$ and $t$. Since addition or deletion of a few edges usually do not change the electrical properties of a network substantially, the above relation makes it easy to see that commute times are also robust to minor perturbation of the graph structure.

We give a very simple but intuitive example. Consider a undirected unweighted graph $G$. We select any node $c$ and add links from $c$ to all other nodes in the graph. Let $i$, and $j$ be any two nodes, s.t. $i \neq j \neq c$. A shortest path between $i$ and $j$ becomes 2 hops now. However the hitting time $h_{ij}$ hardly changes. If a random walk starting at $i$ hits $c$, then there is a very small chance of hitting any other node from $c$, since $\forall j \; p_{c,j} = 1/(n-1)$. Hence the hitting times between the nodes in the graph stay almost the same.

This property is very desirable since in most real-world data-sets there are spurious links resulting from erroneous observations. We also demonstrate this using a simulated dataset in the experimental section.

## 3 Related Work

### 3.1 Applications

The random walks approach has been highly successful in social network analysis (Katz, 1953) and computer vision (Gorelick et al., 2004; Qiu & Hancock, 2005). Here we briefly describe some of these applications.

Saerens et al. (2004) exploit equation (1) to embed a graph and provide distances between any pair of nodes. Yen et al. (2005) replace traditional shortest-path distances between nodes in a graph by hitting and commute times and show that standard clustering algorithms (e.g. K-means) produce much better results when applied to these re-weighted graphs. These techniques exploit the fact that commute times are very robust to noise and provide a finer measure of cluster cohesion than simple use of edge weight.

Gorelick et al. (2004) use the average hitting time of a random walk from an object boundary to characterize object shape from silhouettes. Grady and Schwartz (2006a) introduced a novel graph clustering algorithm which was shown to have an interpretation in terms of random walks. Hitting times from all nodes to a designated node were thresholded to produce partitions with various beneficial theoretical properties. Grady and Schwartz (2006b) used the above approach for automated image segmentation. Also Qiu et al have used commute times clustering for robust multibody motion tracking (Qiu & Hancock, 2006) and image segmentation (Qiu & Hancock, 2005).

Commute times and hitting times have been successfully used for collaborative filtering. In Brand (2005) the authors use different measures resulting from random walks to recommend products to users based on their purchase history . The authors give empirical evidence of the fact that hitting and commute times often are small if one of the nodes has a high degree.whereas the cosine similarity (defined in Section 2) does not have that problem since in some sense the effect of individual popularity is normalized out.

Given a snapshot of a social network, a very interesting question is: which new connections among entities are likely to occur in the future? In Liben-Nowell and Kleinberg (2003) different proximity measures between nodes in a graph are used for link prediction tasks on social networks. The authors showed that the hitting and commute times suffer from the fact that they take into account information from long paths. The most effective measure was shown to be the Katz measure (Katz, 1953) which directly sums over the collection of paths between two nodes with exponentially decaying weights so that small paths are given more weight. However, computing this for all pairs of nodes needs at least $O(n^2)$ time and storage.

### 3.2 Computation

Most clustering applications either require extensive computation to produce pairwise proximity measures, or employ heuristics for restricting the computation to subgraphs. Saerens et al. (2004) compute the trailing eigenvectors of $L$ to reconstruct $L^+$. This still requires at least $O(n^2)$ computation and storage.

Cubic computation is avoided by using sparse matrix manipulation techniques. In Brand (2005) the authors try to compute sub-matrices of $H$ and $C$ by iterative sparse matrix multiplications. Note that computing the $i^{th}$ row of $C$, i.e. $C(i,*)$ or the cosine-similarities of all $j$ with $i$ requires $l_{jj}^+, \forall j$. The exact value of $l_{jj}^+$ requires solving linear systems of equations for all other rows of $L^+$. The authors state that for large markov chains the square of the inverse stationary distribu-tion of $j$ is a close constant factor approximation to $l_{jj}^+$, which is fairly loose. In short, it is only tractable to compute these measures on graphs with a few thousand nodes for most purposes.

The main problem with computing hitting times is that its easy to compute every node's hitting time to a single node, i.e. one column of the hitting time matrix using dynamic programming approaches. However its hard to compute the hitting time from one node to every other node. For that one will have to effectively compute the entries of an entire column in order to get a single entry. This problem can be reduced to the problem of finding all diagonal elements of the pseudo-inverse of the graph Laplacian without computing the entire pseudo-inverse.

### 3.3 Other graph-based proximity measures

Spectral clustering (Ng et al., 2001) is a body of algorithms that cluster datapoints $x_i$ using eigenvectors of a matrix derived from the affinity matrix $W$ constructed from the data. Often $w_{ij}$ is obtained from the Euclidian distance between the datapoints, i.e. $w_{ij} = \exp(\|x_i - x_j\|^2/\sigma^2)$, where $\sigma$ is a free parameter. Zhu et al. (2003) uses graph theoretic approach for semi-supervised learning. Given labeled and unlabeled data-points as nodes in a graph, the main goal is to exploit the graph structure to label the unlabeled examples using the few labeled ones.

## 4 The Algorithm

In this section we present a novel algorithm to compute approximate nearest neighbors in commute time for all nodes, without computing the entire matrix.

### 4.1 Truncated Hitting times

Hitting and commute times decrease if there are many short paths between the nodes. On the other hand, as observed by Liben-Nowell and Kleinberg (2003), two major drawbacks are that they tend to be small whenever one of the nodes has a high degree, and they are also sensitive to parts of the graph far away from the nodes, even when short paths between the nodes exist.

In order to overcome these problems we define a *T-truncated hitting time*, where we only consider paths of length less than $T$. In Section 5 we show that $h^T$ approaches the true hitting time as $T \to \infty$. We use $h$, $h^t$ interchangeably to denote truncated hitting time.

The $T$ truncated hitting time i.e. $h_{ij}^T$ from $i$ to $j$, measures the expected number of steps taken by a random walk starting at node $i$ to hit $j$ for the first time. It can also be defined recursively as

$$h_{ij}^T = 1 + \sum_k p_{ik} h_{kj}^{T-1} \qquad (2)$$

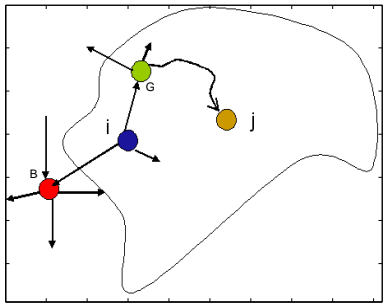

Figure 1: Neighborhood of node $j$. A directed graph is drawn for clarity.

where $h^T$ is defined to be zero if $i = j$ or if $T = 0$. The above equation expresses $h^T$ in a one step look-ahead fashion. The expected time to reach a destination within $T$ timesteps is equivalent to one step plus the average over the hitting times of it's neighbors to the destination. Equation (2) can be easily computed by using Gauss Seidel iterations. Let the truncated hitting times matrix be $H^T$. Note that the hitting time from all nodes to a fixed destination, i.e. a column of $H^T$ can be computed in $O(T\Delta n)$ time, where $\Delta$ is the average degree of a node. However if we need to compute the hitting time from one node to everyone else, i.e. a row of $H^T$, $H^T(i,*)$ then $\forall j$ we will compute the corresponding column in order to get one entry $H^T(i,j)$. Hence we shall end up computing the entire matrix. Instead of examining all pairs of nodes we develop a very efficient framework for doing a range-search to retrieve the $k$ approximate nearest neighbors of *all* nodes in a graph.

### 4.2 GRANCH

The main intuition is that we think about the graph in terms of $n$ overlapping subgraphs. Each subgraph is a bounded neighborhood for one node (discussed in detail later in the section). Consider the hitting times arising from the random walk from any node in this neighborhood to the destination. We provide upper (*pessimistic*) and lower (*optimistic*) bounds of these hitting times, and show how to combine information from all these overlapping subgraphs to obtain $k$ nearest neighbors for all nodes. We will use the terms "optimistic" and "lower", and "pessimistic" and "upper", interchangeably. In order to motivate the GRANCH algorithm we present the following claim before formally describing the algorithm.

**Claim**: Given $k$, $\epsilon$, $T$ and $T' \leq T$, GRANCH returns any $k$ neighbors $j$ of $i$ $\forall i$, s.t. $c^T(i,j) \leq c^T(i, k_{i,T'})(1+\epsilon)$, where $k_{i,T'}$ is the true $k^{th}$ nearest neighbor of $i$ within $T$-truncated hitting time $T'$.

This makes sense since in all applications hitting and commute times are used for ranking entities, e.g. recommend the $k$ best movies based on choices already made by an user; or, find the $k$ most likely co-authors of an author in a co-authorship network.

Suppose we want to estimate $h_{ij}^T$ for pairs of nodes $i,j$ which have relatively low $h_{ij}^T$ values, and suppose we want to avoid $O(n^2)$ time or space complexity. Therefore we cannot do anything that requires representing or iterating over all pairs of nodes. In fact, we cannot even afford to iterate over all pairs of nodes that are less than $T$ hops apart, since in general the number of such pairs is $O(n^2)$. Suppose we restrict computation to iterate only over some subset of pairs of nodes, which we will call the Active-Pairs-set or the $AP$ set. The interesting question is how good are the pessimistic and optimistic bounds we can get on $h_{ij}^T$, $\forall i,j$ using only $O(|AP|)$ work?

As mentioned before we will consider bounded neighborhoods of each node in the graph. For clarity, in case of a directed graph the neighborhood consists of all nodes with short paths *to* the destination node. Here is our algorithm in a nutshell: for each node, we will start with the direct neighbors in its neighborhood, and then compute the optimistic and pessimistic bounds on hitting times of the nodes within a neighborhood to that node. We will then expand the neighborhood and re-iterate. The bounds will be tighter and tighter as we keep expanding the neighborhoods.

Define the set $AP$ as a set of pairs of nodes such that if $i$ is in the neighborhood of $j$, then $(i,j) \in AP$. Note that $(i,j) \in AP$ does not imply that $(j,i) \in AP$. Each neighborhood of node $i$ has a boundary, $\delta_i$, such that a boundary node on the neighborhood of $j$ has direct neighbors outside the neighborhood. Again in a directed graph a boundary node has outgoing edges to nodes outside the neighborhood.

Let's first assume that the neighborhood is given. It's clear from the expression for truncated hitting time that $h_{ij}$ can be computed from the hitting time of $i$'s neighbors to $j$. In Figure 1 the random walk from $i$ can hit either of it's neighbors $G$ or $B$. Since $G$ is inside the neighborhood of $j$ we already have an estimate of $h_{Gj}$. However $B$ is outside the neighborhood of $j$, and hence we find the optimistic and pessimistic values of $h_{Bj}$, resulting into lower bounds ($ho_{ij}$) and upper bounds ($hp_{ij}$) of $h_{ij}$. The optimistic bound is obtained by allowing the random walk to jump to the boundary node with closest optimistic hitting time to $j$, and the pessimistic bound arises from the fact that the walk might never come back to the neighborhood after leaving it, i.e. takes $T$ time.

Now we revisit the data-structures in terms of their notations. We denote the set of active pairs as $AP$.

$AP(*, i)$ is the set of nodes $k$, s.t. $(k, i) \in AP$. Hence the neighborhood of $i$ can also be denoted by $AP(*, i)$. Also $\delta(i)$ is the set of the boundary nodes of the neighborhood of $i$, i.e. $AP(*, i)$. Let $nbs(i)$ denote the set of direct neighbors of $i$. Also denote by $Gnbs(i, j)$ the set $AP(*, j) \cap nbs(i)$. The upper and lower bounds are:

$$hp_{ij}^T = 1 + \sum_{k \in Gnbs(i,j)} p_{ik} hp_{kj}^{T-1}$$
$$+ \left(1 - \sum_{k \in Gnbs(i,j)} p_{ik}\right)(T-1) \quad (3)$$

$$ho_{ij}^T = 1 + \sum_{k \in Gnbs(i,j)} p_{ik} ho_{kj}^{T-1}$$
$$+ \left(1 - \sum_{k \in Gnbs(i,j)} p_{ik}\right)\left(1 + \min_{p \in \delta(j)} ho_{pj}^{T-2}\right) \quad (4)$$

We can also use a tighter bound than the above by using one step lookahead from the nodes on the boundary, which is omitted for space purposes. The $ho$ and $hp$ values of all nodes in $AP(*, j)$ to $j$ can be computed using the Gauss-Seidel technique, as in Algorithm 4.2.

After obtaining the bounds on the hitting times be-

Algorithm 1: compute-H($dst,AP,G,T$)
1: $ho, ho_{last}, hp, hp_{last} \leftarrow ones(1 : N)^1$
2: $minvals_{0:T} \leftarrow 0; minvals_1 = 1$
3: **for** $t = 2$ to $T$ **do**
4:   $min \leftarrow \infty$
5:   **for** $src \in AP(*, dst)$ **do**
6:     $s_1, s_2 \leftarrow 0, prob \leftarrow 0$
7:     **for** $i \in nbs(src)$ **do**
8:       **if** $i \in AP(*, dst)$ **then**
9:         $s_1 \leftarrow s_1 + p_{src,i} ho_{last}(i)$
10:         $s_2 \leftarrow s_2 + p_{src,i} hp_{last}(i)$
11:         $prob \leftarrow prob + p_{src,i}$
12:       **end if**
13:     **end for**
14:     $ho(src) \leftarrow 1 + s_1 + (1 - prob)(1 + minvals_{t-2})$
15:     $hp(src) \leftarrow 1 + s_2 + (1 - prob)(t - 1)$
16:     **if** $ho(src) \leq min$ & $src \in \delta(dst)$ **then**
17:       $min \leftarrow ho(src)$
18:     **end if**
19:   **end for**
20:   $minvals_t \leftarrow min$
21:   $ho_{last} \leftarrow ho$
22:   $hp_{last} \leftarrow hp$
23: **end for**

tween all pairs in the AP-set, we can obtain bounds on hitting times as follows:

$$ho_{ij} = \begin{cases} ho_{ij} & \text{if } i \in AP(*, j) \\ 1 + \min_{p \in \delta(j)} ho_{pj}^{T-1} & \text{else} \end{cases}$$

---
[1] Note that $\forall i, j \ h^1(i, j) = 1$.

$$hp_{ij} = \begin{cases} hp_{ij} & \text{if } i \in AP(*, j) \\ T & \text{else} \end{cases}$$

Note that we are explicitly giving the expressions for all pairs of nodes in the graph for clarity. However the bounds need only $O(|AP|)$ space. Using the above we also get optimistic and pessimistic bounds on commute times, namely $co$ and $cp$.

$$co_{ij} = ho_{ij} + ho_{ji}; \quad cp_{ij} = hp_{ij} + hp_{ji} \quad (5)$$

### 4.3 Expanding Neighborhood

Since time and space complexity of GRANCH is $O(|AP|)$, we have to be clever in having the right set of pairs in the APset, in order to avoid wasting computation on nodes that are very far apart. We present a *neighborhood expansion scheme* which will guarantee that we do not miss any potential nearest neighbors. We also note that adding nodes to the neighborhood of $i$ can only tighten the bounds.

**Theorem 4.1** *The optimistic bounds on H, i.e. ho values can only **increase**, and the pessimistic bounds, i.e. hp values can only **decrease**, as a result of adding nodes to the AP set.*

**Proof**: The hitting time for a node outside the AP-set must be at least the minimum hitting time to the boundary of the AP-set, plus the minimum hitting time from the boundary to the destination. For a destination node $j$, $ho_{m,j}^{T-1} \geq 1 + \min_{p \in \delta(j)} ho_{pj}^{T-2}$, $\forall m \notin AP(*, j)$. Hence $ho_{ij}^T$ in (4) is indeed a valid lower bound on $h_{ij}^T$. Therefore it can never decrease.

The pessimistic bound in (3) is valid, since $h_{ij}^t \leq t$ $\forall i, j$, by construction of $h^t$. Hence the pessimistic bounds can never increase.

**Definition**: Define the lower bound of the neighborhood of node $i$ as $lb(i) = 1 + \min_{p \in \delta(i)} ho_{pi}^{T-1}$.

**Lemma 4.2** *The lower bound on hitting time of nodes outside the AP-set of node $i$, i.e. $lb(i)$, can only **increase** as a result of adding nodes to the AP set.*

**Proof**: This is a direct application of Theorem 4.1.

We start with all direct neighbors of the destination node $i$. Note that all nodes outside the boundary will take at least $1 + \min_{p \in \delta(i)} ho_{pi}^{T-1}$ time to reach $i$. Let us denote this by $lb(i)$. Also remember that we are doing a range search of all neighbors within a range $T' < T$. We want to increase this lower bound, so that we can guarantee that we shall never leave out a potential nearest neighbor. The optimal way to increase the lower bound is to add in the neighbors of $q$, where $q = \arg\min_{p \in \delta(i)} ho_{pi}^{T-1}$.

Algorithm 2: Expand-AP$(G, T, T')$
1: Initialize[2] AP with all pairs $i, j \in E_G$
2: **for** $i = 1$ to $N$ **do**
3:   **while** $lb(i) \leq T'$ **do**
4:     compute-H$(i, AP, G, T)$
5:     $lb(i) \leftarrow 1 + \min_{p \in \delta(i)} ho_{pi}^{T-1}$
6:     $q \leftarrow \arg\min_{p \in \delta(i)} ho_{pi}^{T-1}$
7:     $AP \leftarrow AP \cup \{(a, i) : a \in nbs(q)\}$
8:   **end while**
9: **end for**

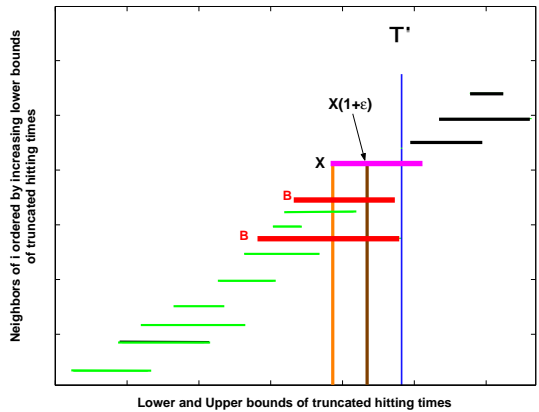

Figure 2: Upper and lower bounds of $h^T(i, *)$

**Lemma 4.3** *If $i \notin AP(*, j)$ after running Algorithm 2, then $h_{ij} \geq T'$.*

After running algorithm 2, $ho_{i,j} \geq lb(j) \geq T'$. Also by construction we have $h_{i,j} \geq ho_{ij}$.

### 4.4 Approximate Nearest Neighbors

We shall first look at how to obtain the $k$ $\epsilon$-approximate nearest neighbors for each node in hitting time. Later we shall demonstrate how to extend that to approximate nearest neighbors in commute times.

Note that we only have upper and lower bounds on hitting times from a node to other nodes. The key idea is that any value $x$ will be smaller than $y$ if the upper bound of $x$ is smaller that the lower bound of $y$.

#### 4.4.1 Truncated hitting time

We illustrate with a toy example. Lets assume that we want to find the $k$ nearest neighbors of node $i$ within $T'$. In Figure 2 we plot the lower and upper bounds of $h_{i,j}$, in the order of increasing lower bounds. It will

---
[2] The AP-set can be initialized to all pairs within $p$ hops too. We used $p = 2$ for some of our experiments.

---

suffice to find all nodes $j$, s.t. $hp_{ij}$ is smaller than the $h$ value of the $(k+1)^{th}$ true nearest neighbor. Note that the $(k+1)^{th}$ nearest neighbor's hitting time is greater than the $(k+1)^{th}$ largest $ho$ value. Hence the upper-bound becomes $X = (k+1)^{th}\_largest\_ho_{i,*}$. Now we also allow a small error margin of $\epsilon$ leading to a new upper bound $X(1+\epsilon)$. All entities with upper bound less than this are guaranteed to be $\epsilon$-approximate.

We also want the neighbors to be within $T'$ hitting time of the source $i$. Hence we find the largest $k' \leq k+1$, s.t. $ho_{i_{k'},i} \leq T'$ and $hp_{i_{k'},i} \geq T'$, where $i_{k'}$ is the node with the $k'^{th}$ largest $ho$ value from $i$. Now the upper bound $X$ becomes $X = ho_{i,i_{k'}}$. We return all nodes $x$, s.t. $hp_{i,x} < X(1+\epsilon)$. Therefore in Figure 2 we have $k = 15$, $k' = 10$. This means that at this point the bounds are not tight enough to determine the exact ranks of the $k - k'$ neighbors. All the nodes with lower bounds beyond $T'$ are guaranteed not to be within $T'$. The nodes labeled $B$ are not guaranteed to be $\epsilon$ approximate and are not returned.

#### 4.4.2 Truncated commute time

Now we look at the neighbors in commute times. The new distance for range-search becomes $2T'$. For any node $i$, we now look at all nodes in the set $AP(i, *) \cup AP(*, i)$. $AP(i, *)$ denotes the set of all nodes that have $i$ in their neighborhoods. After running Algorithm 2, if there exists some node $j$, s.t. $j \notin AP(*, i)$ and $i \notin AP(*, j)$, then by Lemma 4.3, $ho(i, j), ho(j, i) \geq T'$ and hence $co(i, j) \geq 2T'$, so $j$ cannot be within a commute distance of $2T'$ of $i$.

We compute the nearest-neighbors for node $i, i \in \{1 : n\}$ using the APset resulting from Algorithm 2. Now we find the nearest neighbors of $i$ in commute times using equation (5). Note that we only look at the nodes in set $S_i = AP(i, *) \cup AP(*, i)$. Sort all the nodes in increasing order of optimistic bounds i.e. $co$ values. Let $\{i_1, i_2, ...\}$ denote the indexes of the sorted nodes. Let $k'$ the largest integer less than $k$, s.t. $co(i, i_{k'}) \geq 2 * T'$. If such a $k'$ exists define $X$ as $2 * T'$. Otherwise clearly all $k$ possible nearest neighbors might be within the range. Hence define $X$ as $co(i, i_k)$. Return any node $j \in S_i$ s.t. $cp(i, j) \leq X(1+\epsilon)$. Also let $B$ be the set of nodes $b$ in $S_i$ s.t. $co(i, b) \leq X$, but $cp(i, b) \geq X(1+\epsilon)$. Clearly these nodes cannot be guaranteed to be among the $k$ nearest neighbors. Now we expand the AP-sets of $\{i\} \cup B$ to tighten the bounds more.

Because of the construction of the algorithm, the newly introduced nodes can never affect $k'$, and the other nearest neighbors. By Lemma 4.3, we will never introduce a new possible nearest neighbor of a node $b \in \{i\} \cup B$, since all nodes outside $S_b$ are guaranteed to have commute-distance worse than $2T'$. Also an expansion of the AP-set can only increase the lower bounds, and decrease the upper bounds on the estimates on hitting times, which will never decrease the number of neighbors returned so far. Hence we will only improve the performance of the algorithm by expanding our AP-set beyond algorithm 2.

GRANCH uses Algorithm 2 to build the AP-set and then computes the $k$ $\epsilon$-approximate nearest neighbors in truncated commute time.

## 5 Truncated and True Hitting Times

In this section we examine the relation between the true and truncated hitting times. Let $\mathcal{P}$ be the set of paths between $i, j$, such that $j$ appears only once in the path. Let $\mathcal{P}_t$ be all such paths of length exactly $t$. So $\mathcal{P} = \bigcup_t \mathcal{P}_t$. We can then express the truncated hitting time as an expected path length:

$$h_{ij}^T = E(|\mathcal{P}|) = \sum_{t=1}^{T-1} t Pr\{i \xrightarrow{t} j\} + (1 - \sum_t Pr\{i \xrightarrow{t} j\})T$$

Where $Pr\{i \xrightarrow{t} j\}$ denotes the probability that $i$ reaches $j$ for the first time in exactly $t$ steps.

Define $\tilde{P}$ as a matrix identical to $P$, except for the $j^{th}$ row. $\forall m \quad \tilde{P}(j,m) = 0$. Hence after the random walk hits $j$ from $i$, there is a loss of probability mass from the system, which is why the sum is bounded. Thus $Pr\{i \xrightarrow{t} j\} = \tilde{P}_{i,j}^t$. Substituting in the above equation:

$$h_{ij}^T = \sum_{t=1}^{T-1} t \tilde{P}_{ij}^t + (1 - \sum_{t=1}^{T-1} \tilde{P}_{ij}^t)T$$

Note that as $T \rightarrow \infty$ the probability of $i$ not reaching $j$ goes to zero if the graph is connected. Hence the above gets reduced to the true hitting time.

The rate at which the $T$-truncated hitting time approaches the true hitting time depends on the mixing time of a random walk on the graph. The mixing time equals the number of steps before the random walk converges to the stationary distribution. It is the inverse of the mixing rate (Lovasz, 1996) which measures how fast the distribution approaches the stationary distribution. This can be easily computed from the eigenvalues of the graph Laplacian. One way of setting the value of $T$ dynamically will be by evaluating the mixing time. For a graph with a small mixing time, $T$ should be small, since after a few steps the distribution becomes independent of the start node.

## 6 Experiments

We present our experiments in two parts. First we present results on simulated data. Then we give results from large real world data-sets extracted from the Citeseer co-authorship network. We use link-prediction tasks to evaluate the commute times resulting from GRANCH. Link prediction is a very important and fairly hard task in social network analysis. The idea is to predict possible interactions between entities based on the structure of the network.

We present performances of GRANCH and a baseline algorithm, along with the number of pairs of nodes examined by both. The baseline algorithm we use is the number of hops between two nodes, for which we do a simple Breadth First Search from the node upto 4 hops, and define everything beyond as 5 hops. Since social network datasets exhibit transitive social trends, this seemingly naive baseline is often hard to beat. It is used by many online friendship networks for recommending friends. For the small simulated graphs we also computed the exact truncated commute time, which needed all pairs of nodes. Our results show that the performance of GRANCH is comparable to the exact version, even if it looks at small fraction of all pairs of nodes. It also performs better than the baseline, and needs to examine much smaller number of pairs.

Here is a brief description of the experimental methodology. We remove a random subset of links from the original graph. Now we compute the $2T$-truncated-commute times on the training links. The ranking induced by this is tested against the held out links using their AUC-scores. We only take nodes which has at least one link held out. For each such node we take the proximity measures from the algorithms of all nodes within 4 hops of the source in the original graph(not all other nodes) and compute the AUC score of that vector. We present the average AUC score.

We denote the exact commute time by $C$. We also compare our algorithm for different values of $T$, $3, 6, 10$. For most of the graphs we noticed that $T = 6$ was enough. $T$ in all tables translates to a $2T$-truncated commute time. We also saw that as we increase our range of search, i.e. $T'$ we get closer to the accuracies of the exact truncated hitting times. A typical choice of $T'$ is 2.9 for $T = 3$, 5.95 for $T = 6$, and 9.75 for $T = 10$.

We present the total number of pairs in the $AP$-set for GRANCH in Tables 2 and 4. These tables also show the total number of pairs the baseline needs to look at while examining all nodes with at least one edge held out. This is because we compute AUC scores for only these nodes. We also put the total number of pairs within 4 hops for the Citeseer graphs in Table 5. The total number of pairs for the baseline can be as much as these numbers in the worst case.

### 6.1 Simulated Data

A graph $G$ can be quantified in terms of its growth rate $\rho_G$ (Krauthgamer & Lee, 2003). $\rho_G$ is defined as the minimum $\rho$, s.t. every ball of radius $r > 1$ in $G$ contains at most $O(r^\rho)$ nodes. By definition, this implies that the total number of nodes within $c$ hops of any node is at most $O(c^{\rho_G})$. For example the 2D grid graph has $\rho = 2$, and in general a $k$ dimensional hypergrid has $\rho = k$.

Table 1: AUC score on 30% links held out from graphs of size $n$, edges $e$, growth-rate $d$ for algorithms $C$, $GRANCH$ with $T = 10$, $T = 6$, $T = 3$ and the baseline (number of hops between node-pairs within 4 hops).

| $d$ | $n$ | $e$ | C,T=10 | GRANCH,T=10 | C,T=6 | GRANCH,T=6 | C,T=3 | GRANCH,T=3 | Baseline |
|---|---|---|---|---|---|---|---|---|---|
| 2 | 1,000 | 2700 | 82.4 | 82.4 | 82.4 | 82 | 71 | 71 | 69 |
| 4 | 1,000 | 2700 | 85 | 83 | 85 | 83 | 68 | 68 | 74 |
| 2 | 10,000 | 27,000 | - | 82.2 | - | 83 | - | 73 | 70.6 |
| 4 | 10,000 | 27,000 | - | 80 | - | 80 | - | 66 | 73 |

Table 2: On training set of 70% links from simulated graphs of dimension $d$, size $n$, edges $e$, number of pairs in AP-set of GRANCH and number of pairs examined by algorithm $C$ (with $T = \{10, 6, 3\}$) and the baseline.

| $d$ | $n$ | $e$ | C,T=10 | GRANCH,T=10 | C,T=6 | GRANCH,T=6 | C,T=3 | GRANCH,T=3 | Baseline |
|---|---|---|---|---|---|---|---|---|---|
| 2 | 1,000 | 2,700 | 1M | 24,594 | 1M | 29,656 | 1M | 13,521 | 34,902 |
| 4 | 1,000 | 2,700 | 1M | 47,846 | 1M | 45,623 | 1M | 14,839 | 72,646 |
| 2 | 10,000 | 27,000 | - | 257,110 | - | 311,944 | - | 141,136 | 381,564 |
| 4 | 10,000 | 27,000 | - | 503,998 | - | 478,409 | - | 170,144 | 781,515 |

We generate graphs of growth-rate $d$, by assigning $d$ dimensional Euclidian coordinates to the nodes, and adding links between close neighbors plus some long distance links. This brings about the very famous *small world phenomenon* (Milgram, 1967). We do the following experiments,

1. AUC scores on held out data
2. AUC scores on data with noisy links

We only present the performance of the exact truncated commute time for the 1,000-node graph, since the computation becomes too expensive for the 10,000-node graph. As in Table 1, our algorithm beats the baseline algorithm, and does almost as well as the exact truncated commute time(for the 1000 node graph) without looking at all pairs of nodes.

Also from Table 2 we can see how for a given number of nodes and edges the dimensionality of the graph affects the number of pairs needed by both GRANCH and the baseline algorithm.

**Noisy Data**: We generate an example where we make a dataset noisy by adding links from a randomly chosen node to 20% of the remaining nodes. Now we hold out 30% links from this graph, and test the AUC score of our algorithm. We expect to see that the commute times outperform the baseline, since it's much more robust to noise. For a 100-node graph, GRANCH with $T = 6$ has an AUC score of 85.3%, whereas the number of hops has an AUC score of 70.23%.

### 6.2 Real world graphs

In this section we perform similar experiments as in the last section on real world graphs. We use subsets of the Citeseer network of sizes: 20,000 nodes with 55,000 edges, 50,000 nodes with 160,000 edges and 100,000 nodes with 280,000 edges. These graphs are connected but sparse. We present the total number of pairs within 4 hops of one another in the training graphs in table 5. These numbers reflect the small world phenomenon in real-world social networks. Every node can reach every other node within a few hops. This also emphasizes the fact that its not even feasible to take all nodes within $T$ hops to compute the truncated commute times exactly. For reasonable values of $T$, the above will be a huge fraction of all pairs.

Table 5: Number of pairs within 4 hops in training graphs(90% links from Citeseer) of different sizes.

| $n$ | $e$ | # of pairs within 4 hops |
|---|---|---|
| 19,477 | 54,407 | 3,401,812 |
| 52,033 | 152,518 | 20,478,196 |
| 103,500 | 279,435 | 88,418,488 |

We remove a randomly chosen 10% of the links to obtain the training and test graphs, with the same setting of $T'$. For these datasets, the exact truncated commute time is intractable. In Table 4 note that the baseline has to compute hops between number of pairs that is more that one order of magnitude larger than the number of pairs considered by GRANCH. Also the numbers in Table 4 are only for the nodes which had at least one edge held out from the original graph. In the worst case these numbers could be as large as in Table 5. The baseline algorithm performs quite well for these datasets indicating existence of transitive trends in co-authorship networks.

Table 3: AUC score of GRANCH and the baseline on 10% links held out from graphs of size $n$, edges $e$ with $T = 10$; $T = 6$; $T = 3$

| $n$ | $e$ | GRANCH, $T=10$ | GRANCH, $T=6$ | GRANCH, $T=3$ | Baseline |
|---|---|---|---|---|---|
| 19477 | 54407 | 82 | 80 | 78 | 74 |
| 52033 | 152518 | 85 | 84 | 78.6 | 79 |
| 103500 | 279435 | 86.7 | 86 | 79 | 83 |

Table 4: On the training set of 90% links from graphs of size $n$, edges $e$, number of pairs in AP-set of GRANCH with $T = 10$; $T = 6$; $T = 3$; and number of node-pairs used by the baseline.

| $n$ | $e$ | GRANCH, $T=10$ | GRANCH, $T=6$ | GRANCH, $T=3$ | Baseline |
|---|---|---|---|---|---|
| 19,477 | 54,407 | 600,000 | 335,769 | 279,736 | 1,336,683 |
| 52,033 | 152,518 | 2,017,931 | 1,532,534 | 293,366 | 8,723,436 |
| 103,500 | 279,435 | 5,143,982 | 4,156,792 | 1,661,598 | 40,563,881 |

## 7 Conclusion and Future Work

This paper has addressed an important gap in graph proximity measures: that the important measures are also the hardest to compute. We designed an algorithm that guarantees an $\epsilon$-error and showed empirically that it avoids $O(n^2)$ computation or storage in practice while retaining the predictive performance of the expensive commute time methods.

For future directions we will like to analyze the runtime of GRANCH. We would also like to do a detailed comparison of our algorithm with different link prediction algorithms. Currently we set the parameter $T$ for truncation at different values to have different behavior of the algorithm. It would be interesting to choose $T$ dynamically based on the mixing time of the graph.

## 8 Acknowledgements

We are very grateful to Jeremy Kubica for his helpful comments and suggestions. This work was done while the first author was an intern at Google Pittsburgh.